\documentclass[12pt,twoside,a4paper,amsmath,amssymb,apsmy,showkeys,showpacs,nofootinbib]{revtex4}

\usepackage[cp1251]{inputenc}
\usepackage[russian,english]{babel}
\usepackage[left=2.5cm, right=2.5cm, top=2.5cm, bottom=2.2cm, bindingoffset=0cm]{geometry}

\usepackage[T2A]{fontenc}

\usepackage{subfig}
\usepackage{graphicx}
\usepackage{graphics}
\usepackage{epsfig}
\usepackage{float}
\usepackage{multirow}
\usepackage{bigstrut}
\usepackage{latexsym}
\usepackage{array}
\usepackage{dcolumn}
\usepackage{bm}

\begin{document}

\thispagestyle{myheadings}

\title{Phase diffusion in high-power microwave sources}

\author{Anishchenko S.V.}
\email{sanishchenko@mail.ru}
\affiliation{Research Institute for Nuclear Problems of Belarusian State University \\Bobruiskaya Str. 11, 220030 Minsk, Belarus}

\author{Baryshevsky V.G.}
\email{bar@inp.bsu.by}
\affiliation{Research Institute for Nuclear Problems of Belarusian State University \\Bobruiskaya Str. 11, 220030 Minsk, Belarus}

\author{Gurinovich A.A.}
\email{gur@inp.bsu.by}
\affiliation{Research Institute for Nuclear Problems of Belarusian State University \\Bobruiskaya Str. 11, 220030 Minsk, Belarus}

\begin{abstract}
	Autocorrelation of electromagnetic fields emitted by high-power microwave sources makes it possible to determine the phase diffusion coefficient $D$. The value of $D$ imposes significant constraints on synchronization of several HPM sources. It is clearly demonstrated by the example of a relativistic reflex triode operating at $f\approx3.3$~GHz and having the phase diffusion coefficient equal to $D\approx0.06$~ rad$^{2}\cdot$ns$^{-1}$.
\end{abstract}


\maketitle

\section{Introduction}
In low-current electronics, the nature of noise was established more than sixty years ago~\cite{Schottky1918,Johnson1925,Johnson1928,Nyquist1928,Bernamont1937,Burgess1954}. Investigation of noise and fluctuations made it possible to understand spectral lines broadening, amplitude fluctuations, and phase diffusion in physical oscillators~\cite{Pontryagin1933,Bernshtein1938,Bernstein1941,Bernstein1950}, study oscillators synchronization in detail ~\cite{Stratonovich1961,Malakhov1967}, and create modern radars based on phased antenna arrays.

In relativistic high-power microwave (HPM) sources, noise is not studied so well. Despite the fact that the first high-current oscillators appeared in the 70s of the last century~\cite{Kovalev1973,Carmel1974,Bugaev1979}, only a few works on the topic have been published to date~\cite{Abubakirov2009,Abubakirov2018,Shunailov2019}. In accordance with the published material, the main source of noise in HPM sources is a complex structure of electron beams produced due to explosive electron emission~\cite{Mesyats1971,Bugaev1975}. A high-current electron beam consists of many portions of electrons. The portions are called ectons. Each ecton contains up to $10^{11}$ of elementary charges~\cite{Bugaev1979, Bugaev1975}. It is obvious that ecton structure of e-beams affects the statistical properties of the radiation emitted by high-current e-beams~\cite{Abubakirov2009} and synchronization of several HPM sources.

Publications on the synchronization of HPM sources have been appearing in the scientific press since the second half of the 1980s~\cite{Woo1989,Sze1990,Benford1989,Levine1991,Kanaev1995,Vintizenko2013,Ju2016}. Despite the detailed description of various connection topologies between generators, any constraints on noise properties for stable operation of coupled systems were not revealed in the mentioned studies.

In this regard, the paper presents an approach to the experimental study of noise in HPM sources. The study is based on the determination of the phase diffusion coefficient $D$ by means of radiation field autocorrelation. The fruitfulness of the approach is demonstrated by the example of a relativistic reflex triode operating at 3.3~GHz~\cite{Baryshevsky2017}.
We will also describe the limitations that the phase diffusion imposes on synchronization of microwave oscillators. The results can be used in the design of HPM phased arrays~\cite{Baryshevsky2019}.

\section{Autocorrelation}
HPM sources are characterized by short operation time. As a rule, the stationary operation time $t_{s}$ is comparable with the rise time $t_{f}$. Therefore, the use of autocorrelation for the analysis of short pulses requires special consideration.

\begin{figure}[ht]
	\leavevmode
	\centering
	\resizebox{160mm}{!}{\includegraphics{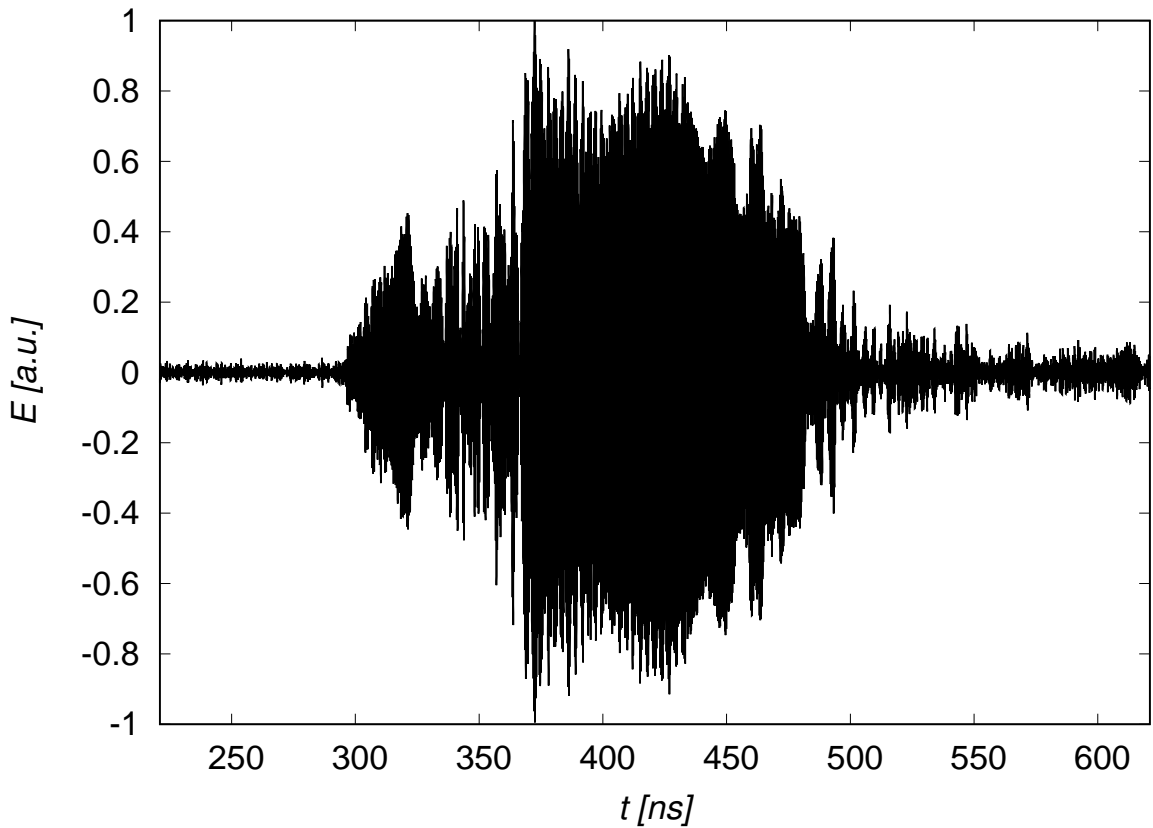}}\\
	\caption{Microwave pulse.}
	\label{fig:experiment}
\end{figure}

\begin{figure}[ht]
	\leavevmode
	\centering
	\resizebox{160mm}{!}{\includegraphics{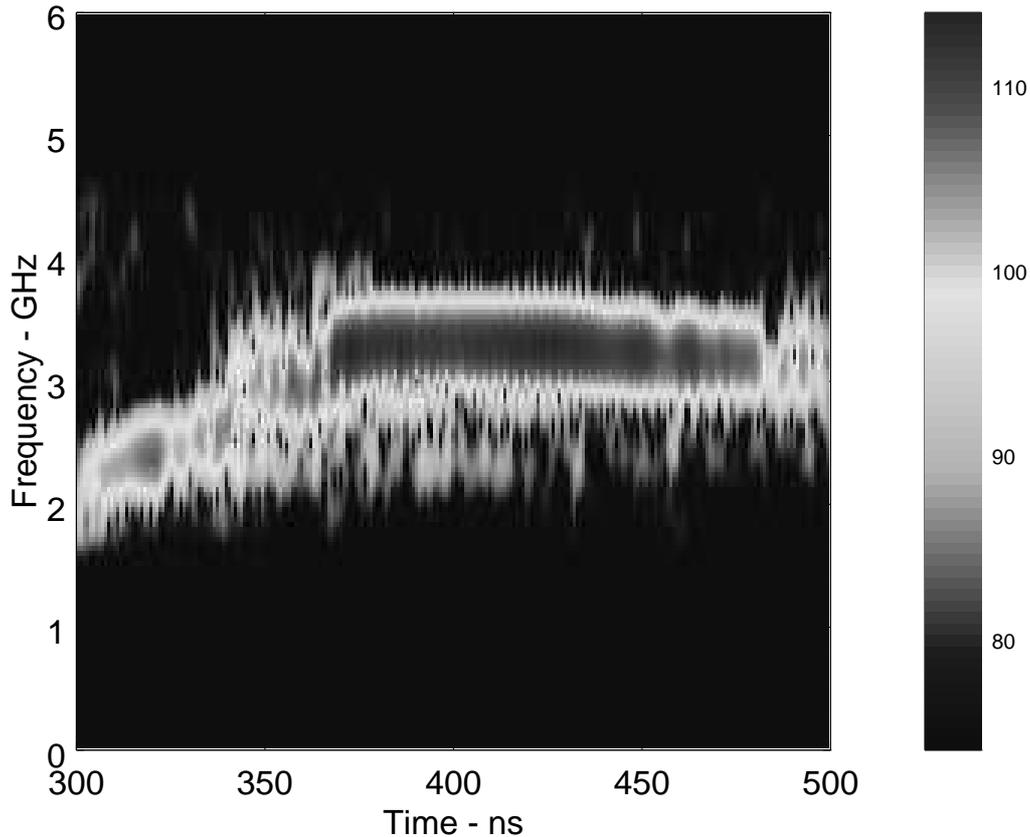}}\\
	\caption{Wavelet transform of measured electric field.}
	\label{fig:wavelet}
\end{figure}

\begin{figure}[ht]
	\leavevmode
	\centering
	\resizebox{160mm}{!}{\includegraphics{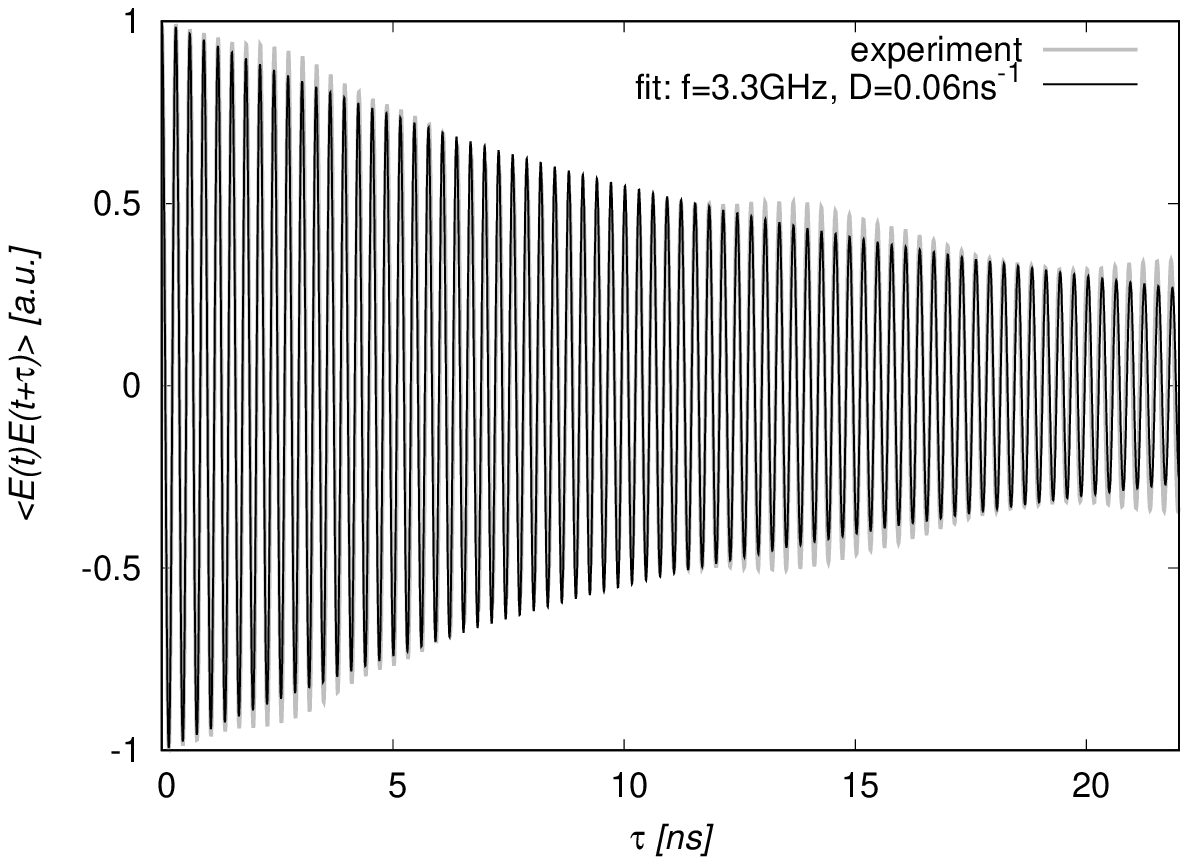}}\\
	\caption{Normalized autocorrelation function.}
	\label{fig:autocorrelation}
\end{figure}

In addition, it should be taken into account that HPM sources are complex distributed multimode systems. As a result, generation process is carried out on several modes. In this regard, special measures should be taken to enhance the main operation mode and suppress all the others.
The operation mode is usually chosen based on the voltage <<flat top>> applied to an HPM source. Meanwhile, the generation conditions change significantly at the voltage rise and drop: the synchronism condition between the high-current electron beam and the operation mode is violated. As a result, mode hoping is observed.
In the experiment, mode hoping manifests itself through new frequency lines in the emission spectrum. Their presence imposes additional limitations on noise study using autocorrelation that is well-suited for single-frequency signals~\cite{Pikovsky2004}.

Thus, we can conclude that noise analysis for short-pulse signals requires at least refinement. In this paper, when analyzing noise, we restrict ourselves to the analysis of signal time frames corresponding to stationary generation.

During stationary generation, the phase dynamics is described by the stochastic differential equation
\begin{equation}
\label{eq:dPhi}
\frac{d\Phi}{dt}=\omega+\xi(t),
\end{equation}
where $\xi(t)$ is a white-noise function satisfying the condition
\begin{equation}
<\xi(t_1)\xi(t_2)>=2D\delta(t_2-t_1).
\end{equation}
Here, $\delta(t)$ is the Dirac function.

The solution of the equation \eqref{eq:dPhi} has the form
\begin{equation}
\label{eq:Phi}
\Phi(t)=\phi_0+\omega t+\phi(t)
\end{equation}
where $\phi(t)=\int_0^t\xi(\tau)d\tau$ is a random phase shift with the the mean value 0 and variance
\begin{equation}
\label{eq:dispersion}
<\phi^2(t)>=2Dt.
\end{equation}
The distribution function for the random phase shift $\phi$ satisfies the diffusion equation
\begin{equation}
\frac{\partial G}{\partial t}=D\frac{\partial^2G}{\partial\phi^2},
\end{equation}
whose solution is the well-known gaussian distribution
\begin{equation}
G(\phi,t)=\frac{1}{\sqrt{4\pi Dt}}\exp\Big(-\frac{\phi^2}{4Dt}\Big).
\end{equation}

Let $E(t)=E_0\cos\big(\Phi(t)\big)$ be an oscillating electric field with a random phase $\Phi(t)$  (see \eqref{eq:Phi}).
Then, averaging the product of electric fields $E(t)E(t+\tau)$ using $G(\phi,\tau)$, we calculate the autocorrelation\footnote{We use the generally used definition for the diffusion coefficient~\cite{Kampen1990}, which is different from~\cite{Pikovsky2004}. As a consequence, the variance of the random phase shift is equal to $2Dt$.}
\begin{equation}
\label{eq:ac}
K(\tau)=\frac{1}{t_s}\int_0^{t_s}<E(t)E(t+\tau)>dt=\frac{E_0^2e^{-D\tau}\cos(\omega\tau)}{2}.
\end{equation}

Expression \eqref{eq:ac} shows that $K(\tau)$ is a damped function oscillating with frequency $\omega$. The damping constant $K(\tau)$ is equal to the diffusion coefficient $D$. This means that $D$ can be determined from the autocorrelation of measured signals.

Now, by means of a relativistic reflex triode, we will demonstrate how the autocorrelation \eqref{eq:ac} can be used to determine the phase diffusion coefficient $D$.
The figure~\ref{fig:experiment} shows the experimental curve for measured electric field~$E(t)$. The wavelet transform of~$E(t)$ shows (figure~\ref{fig:wavelet}) that the stationary single-frequency operation lasts $t_s\approx80$~ns (from 370ns to 450~ns). Lower frequencies are observed at the pulse rise and drop.

For the stationary operation, we plot autocorrelation and its fit of the form \eqref{eq:ac} (Fig.~\ref{fig:autocorrelation}).
The diffusion coefficient, which determines the damping constant for $K(t)$, turns out to be equal to $D\approx0.06$~ns$^{-1}$.
Known $D$ allows one to investigate synchronization of several HPM sources.

\section{Synchronization}
Synchronization between two reflex triodes can be investigated with the help of the Langevin equations describing phase dynamics
\begin{equation}
\label{eq:langevin}
\begin{split}
&\dot\Phi_1=\omega+\varepsilon\sin(\Phi_2(t-\tau)-\Phi_1(t))+\xi_1(t),\\
&\dot\Phi_2=\omega+\varepsilon\sin(\Phi_1(t-\tau)-\Phi_2(t))+\xi_2(t).\\
\end{split}
\end{equation}
Here, $<\xi_1(t_1)\xi_1(t_2)>=2D\delta(t_1-t_2)$ and $<\xi_2(t_1)\xi_2(t_2)>=2D\delta(t_2-t_1)$ are standard autocorrelations for white noise. Positive quantities $\varepsilon$ and $\tau$ represent coupling coefficient and time delay between HPM sources, respectively.

To make radiation sources to oscillate in-phase, $\omega\tau$ should be equal to an integer number of $2\pi$~\cite{Woo1989}. If we assume that $\varepsilon\tau\ll1$ then the standard procedure for solving stochastic delay differential equations can be applied \cite{Guillouzic1999}. The procedure is based on the following expansion of time delayed functions: $\Phi_{1,2}(t-\tau)\approx\Phi_{1,2}(t)-\tau\dot{\Phi}_{1,2}(t)$.
As a result, the Langevin equations are reduced to a single equation for the phase difference $\psi=\Phi_2-\Phi_1$:

\begin{equation}
\frac{d\psi}{dt}=-\frac{2\varepsilon}{1-\varepsilon\tau}\sin(\psi)+\xi_{12}(t),
\end{equation}
where $\xi_{12}(t)=\frac{\xi_2(t)-\xi_1(t)}{1-\varepsilon\tau}$ is a random white-noise function with autocorrelation
\begin{equation}
<\xi_{12}(t_1)\xi_{12}(t_2)>=\frac{4D}{(1-\varepsilon\tau)^2}\cdot\delta(t_2-t_1).
\end{equation}


The distribution function $P$ for the phase difference $\psi$ obeys the Fokker-Planck kinetic equation
\begin{equation}
\label{kinetic_equation}
\frac{\partial P}{\partial t}=\frac{2\varepsilon}{1-\epsilon\tau}\frac{\partial}{\partial\psi}\big(\sin(\psi)P\big)+\frac{2D}{(1-\varepsilon\tau)^2}\frac{\partial^2P}{\partial\psi^2}.
\end{equation}
Note that the diffusion coefficient~$2D$ for the phase difference~$\psi$ is more than two times greater than the diffusion coefficient~$D$ in a single oscillator. The equation \eqref{kinetic_equation} differs from standard Fokker-Plank equation without time delay~\cite{Stratonovich1961,Malakhov1967,Pikovsky2004} by coefficients~$\frac{1}{1-\varepsilon\tau}$ and $(\frac{1}{1-\varepsilon\tau})^2$.

\begin{figure}[ht]
	\leavevmode
	\centering
	\resizebox{80mm}{!}{\includegraphics{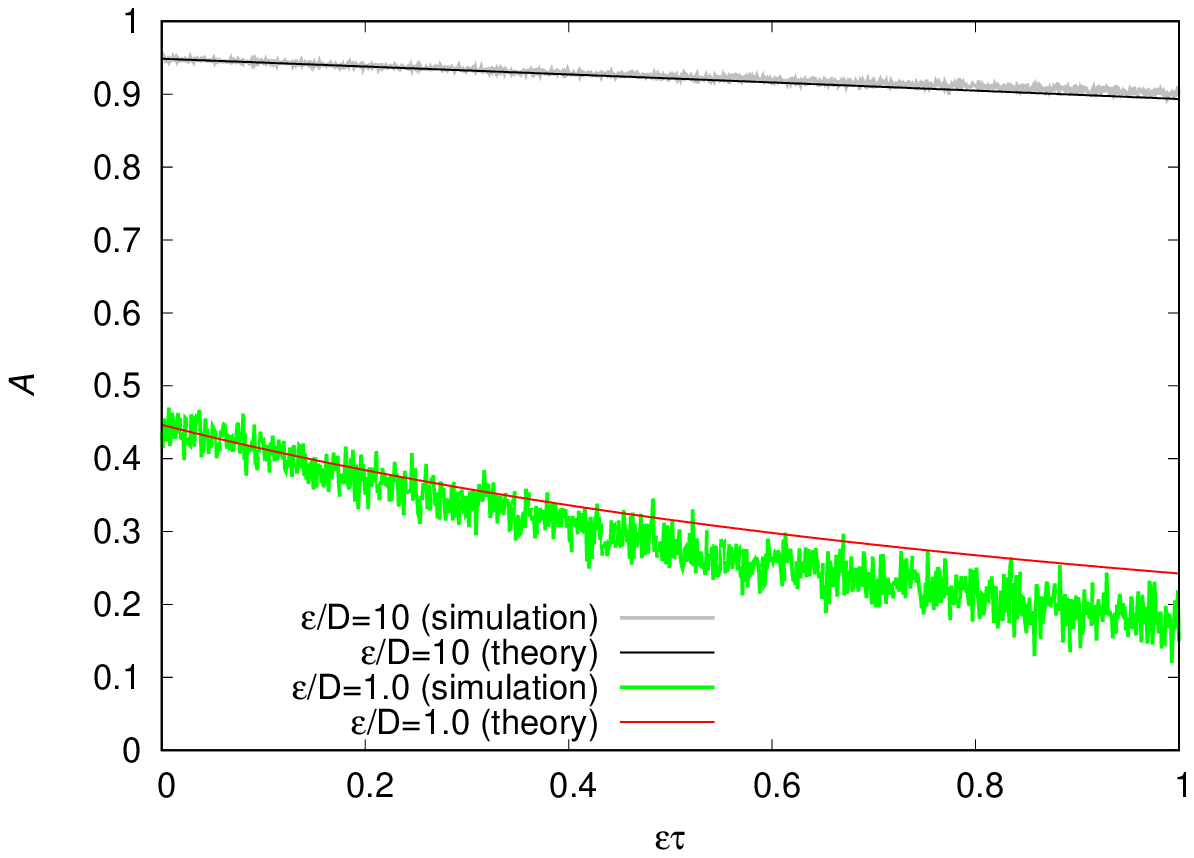}}\resizebox{80mm}{!}{\includegraphics{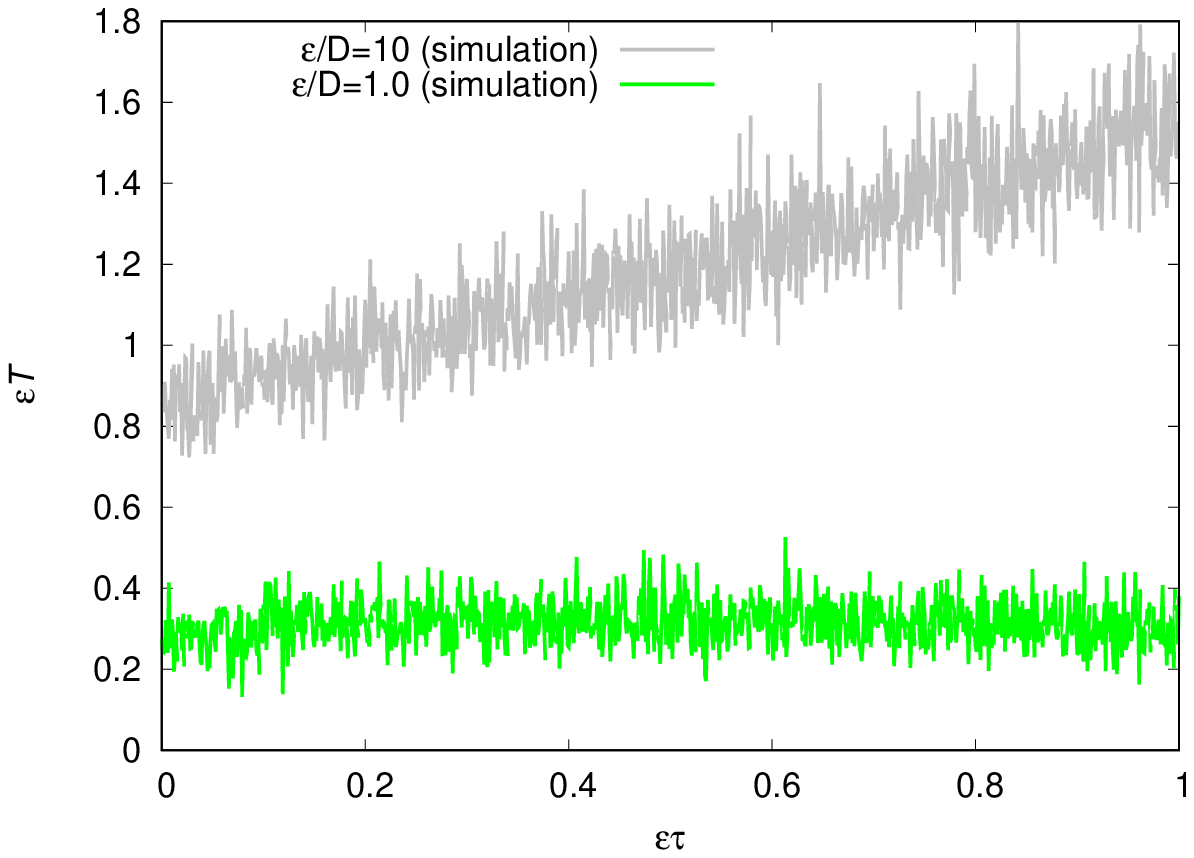}}\\
	\caption{Mean value $A$ (left) and average synchronization time $T$ (right).}
	\label{fig:cos}
\end{figure}

For stationary operation, the solution of \eqref{kinetic_equation} is the time-independent Stratonovich distribution
\begin{equation}
\label{eq:stratonovich}
P(\psi)=\frac{\exp\big(\alpha\cos(\psi)\big)}{2\pi I_0(\alpha)},
\end{equation}
where  $I_0$ and 
\begin{equation}
\label{eq:alpha1}
\alpha=\varepsilon(1-\varepsilon\tau)/D
\end{equation}
are the modified Bessel function and the dimensionless parameter, which is responsible for synchronization, respectively.

The average value $<\cos(\Phi_2-\Phi_1)>=<\cos(\psi)>$, which we denote by $A$, determines the average radiation intensity of two HPM sources $J_2=2J_1\big(1+A\big)$ ($J_1$ is radiation intensity of a single source). The quantity $A$ can be calculated using the following formula
\begin{equation}
A=<\cos(\psi)>=\int_{-\pi}^{+\pi}P(\psi)\cos(\psi)d\psi=\frac{I_1(\alpha)}{I_0(\alpha)}.
\end{equation}
For $\alpha\gg1$, the behavior of $A$ is approximately described by the relation
\begin{equation}
\label{eq:asymptotics}
A\approx 1-\frac{1}{2\alpha}.
\end{equation}

Figure~\ref{fig:cos} shows the decrease in $A$ with growing $\varepsilon\tau$ for different $\varepsilon/D$.
(When constructing theoretical curves, we resorted to one trick. Namely, we replace \eqref{eq:alpha1}, which has unphysical behavior at $\epsilon\tau\sim1$, by
\begin{equation}
\label{eq:alpha2}
\alpha=\frac{\varepsilon}{D(1+\varepsilon\tau)}.
\end{equation}
Both approximations have the same behavior at $\varepsilon\tau\ll1$. But the latter is positive and finite for all time delays.)
To enhance synchronization, it is necessary to increase the coupling coefficient $\varepsilon$ and decrease the time delay $\tau$ and the diffusion coefficient $D$.
For $\varepsilon=0.6$~ns$^{-1}$, $D=0.06$~rad$^2\cdot$ns$^{-1}$, and $\varepsilon\tau=1$, we get $A\approx0.95$ which is 5\% less than the maximum possible value ($A\to1$ for $\alpha\to+\infty$).

Let us pay attention to the average value of the synchronization time $T$. (We assume that synchronization time is the time $t$, when the random variable $\cos\big(\psi(t)\big) $ becomes equal to $A$.) The quantity $T$ can be obtained from numerical solution of the stochastic Langevin equations \eqref{eq:langevin} (see Fig.~\eqref{fig:cos}). Let $\varepsilon=0.6$~ns$^{-1}$,~$D=0.06$~rad$^{2}\cdot$ns$^{-1}$, and $\varepsilon\tau=1$, then  $T=0.98/\varepsilon\approx2$~ns is much less than stationary operation time ($t_s\approx80$~ns). As a result, synchronization of two reflex triodes seems to be plausible.

\section{Conclusion}
Autocorrelation of electromagnetic fields emitted by HPM sources makes it possible to determine the phase diffusion coefficient $D$. The numerical value of $D$ imposes significant limitations on synchronization of several HPM sources. This is demonstrated by means of a relativistic reflex triode operating at $f\approx3.3$~GHz and having phase diffusion coefficient $D\approx0.06$~rad $^{2}\cdot$ns$^{-1}$. With a coupling coefficient equal to 0.6~ns$^{-1}$, synchronization of two relativistic reflex triodes should occur in 2~ns.

\label{last}
\end{document}